\documentclass[prx,longbibliography,reprint,showpacs,amsmath,amssymb,groupedaddress,superscriptaddress]{revtex4-1}
\usepackage{graphicx,color}
\usepackage{dcolumn}
\usepackage{float}
\usepackage{multirow,tabularx,booktabs} 

\begin{document}
    
    \title{ Properties of real metallic surfaces: Effects of density functional semilocality and van der Waals nonlocality  }
    
    \author{Abhirup Patra}
    \affiliation{Department of Physics,Temple University, Philadelphia,PA-19122}
    
    \author{Jefferson E. Bates}
    \affiliation{Department of Physics,Temple University, Philadelphia,PA-19122}
    \author{Jianwei Sun}
    \affiliation{Department Of Physics, University of Texas-El Paso, El Paso,TX-79902}
    \author{John P. Perdew}
    \affiliation{Department of Physics \& Chemistry ,Temple University, Philadelphia,PA-19122}
    \date{\today}
    
\begin{abstract}
    We have computed the surface energies, work functions, and interlayer surface relaxations
    of clean (111), (110), and (100) surfaces of Al, Cu, Ru, Rh, Pd, Ag, Pt, and Au.
    Many of these metallic surfaces have technological or catalytic applications. We
    compare experimental reference values to those of a family of non-empirical semilocal
    density functionals from the basic local density approximation (LDA) to our most
    advanced, general-purpose meta-generalized gradient approximation, SCAN.  The closest
    agreement within experimental uncertainty is achieved by the simplest density functional
    LDA, and by the most sophisticated general-purpose one, SCAN+rVV10. The long-range
    van der Waals interaction incorporated through rVV10 increases the surface energies
    by about 10\%, and the work functions by about 3\%. LDA works for metal surfaces
    through a stronger-than-usual error cancellation. The PBE generalized gradient
    approximation tends to underestimate both surface energies and work functions,
    yielding the least accurate results. Interlayer relaxations from
    different functionals are in reasonable agreement with one another, and usually
    with experiment.
\end{abstract}
    
    \maketitle

\section{\label{sec:level1}Introduction}
    The rapid development of electronic structure theory has made it easier to analyze
    and describe complex metallic surfaces\cite{gross2014theoretical}, but understanding the underlying physics 
    behind surface energies, work functions, and interlayer relaxations has remained 
    a long-standing challenge\cite{DaSilva}. Metallic surfaces are of particular importance
    because of their wide range of applications including metal-molecule junctions, 
    junction field-effect transistors, and 
    in catalysis \cite{MRS-JOHN,MRS-WHITE,catalysis-1,catalysis-2,catalysis-3,catalysis-4}.
    A detailed knowledge of the electronic structure is required for accurate theoretical investigations 
    of metallic surfaces \cite{SurfaceInglesfield,Sahnirecent}.
    
    Consequently, metal surfaces have played a key role in the development and application of Kohn-Sham
    density functional theory (KS DFT) \cite{KOHN-DFT2}. The work of Lang and Kohn \cite{KOHN-SURFACE,KOHN-WORK,KOHN-IMAGE}
    in the early 1970's demonstrated the ability of the simple local density approximation (LDA) \cite{KOHN-DFT2,PW92} 
    for the exchange-correlation (xc) energy to capture the surface energies
    and work functions of real metals. Their work stimulated the effort to understand why
    simple approximate functionals work and how they can be improved \cite{LP77,LP80}. Later,
    correlated-wavefunction calculations \cite{KK85,AC96} gave much higher surface energies for
    jellium, but were not supported by further studies \cite{AP02,CP08} and were eventually
    corrected by a painstaking Quantum Monte Carlo calculation \cite{WH07}. The too-low surface
    energies from the Perdew-Burke-Ernzerhof \cite{PBE} (PBE)  generalized-gradient approximation
    (GGA) led in part to the AM05 \cite{AM05}  and PBEsol \cite{perdew2008restoring} (PBE for solids) GGAs,
    and to general-purpose meta-GGAs that remain computationally efficient, including
    the recent strongly constrained and appropriately normed (SCAN) meta-GGA \cite{SCAN, SR16}.
    SCAN captures intermediate-range van der Waals (vdW) interactions, but capturing 
    longer-ranged vdW interactions requires the addition of a non-local vdW correction such as from
    the revised Vydrov-Van Voorhis 2010 (rVV10) functional \cite{SG13}. 
    
    Ref.~\onlinecite{rVV10} suggests that the vdW interaction is semilocal at short and intermediate range, but displays 
    pairwise full nonlocality at longer ranges, and many-body full nonlocality \cite{AF16}
    at the longest and least energetically important distances. Accounting for intermediate and long-ranged vdW interactions is especially important for 
    layered materials \cite{LHG10,BGK12,SCAN+rVV10} and ionic solids \cite{ZGU13,TZG17}.
    van der Waals interactions are also needed to correct the errors of GGAs for bulk metallic systems \cite{TZG17}. 
    The importance of the vdW contribution to surface properties is something we emphasize below.
    By naturally accounting for both intermediate and long-range interactions,
    SCAN+rVV10 \cite{rVV10} represents a major improvement over previous functionals for many properties of diversely-bonded
    systems \cite{SR16}, however, it had not been tested for real metallic surfaces. 
    By studying metallic surfaces with this general-purpose functional we can better 
    understand why LDA can be accidentally accurate, and demonstrate the systematic  
    improvement of SCAN over other non-empirical functionals. 
    Furthermore, we will also be able to extract the impact of intermediate and long-range dispersion.
    
    The surface energy is the amount of energy required per unit area to cleave an
    infinite crystal and create a new surface \cite{KOHN-SURFACE}. Accurate theoretical face-dependent
    surface energies are straightforward to obtain from accurate bulk and surface calculations,
    since we have absolute control over morphology and purity. Experimentally, however, 
    surface energies have been determined by measuring the surface tension of the liquid
    metal and then extrapolating to 0K using a phenomenological method \cite{al111surfexp}. The surface
    tension of the liquid phase is generally different from the actual surface free 
    energy of the solid state metals and can be considered as an ``average'' surface energy.
    Available experimental values are also rather old
    (1970-1980). They provide useful but uncertain estimates for the low-energy faces
    of bulk crystals. With this in mind, one should be careful when comparing defect-free
    theoretical predictions with experimental data. 
    
    The work function on the other hand, is easier to measure \cite{Rohwerder2007290} compared to the surface energy.
    One can consider the work functions measured from a polycrystalline sample \cite{BOOK} as a ``mean'' work function
    of all the possible crystallographic faces present in the sample. 
    In a recent work \cite{Helander-work}, 
    Halender et. al argued that ultraviolet photoemission spectroscopy can be used as an accurate measurement of the work function of materials, 
    despite of the chances of surface contamination of the exposed surface. 
    However, it still remains an open question as to which experimental work function value one should compare theoretical values with.
    In practice the work function can be calculated
    using DFT by accurately determining the Fermi energy and vacuum potential \cite{KOHN-WORK, KOHN-IMAGE}
    of the surface slab. Since the work function depends on slab thickness, the role
    of quantum size-effects may need to be considered. 
    
    Another fundamental property of surfaces is geometric relaxations due to strong
    inward electrostatic attraction on the top surface layer, which occurs mainly due 
    to charge density smoothing at the surface. This effect can be accurately measured
    experimentally using low-energy electron diffraction (LEED) \cite{al111surfexp, alLEED}  intensity
    analysis. The role played by the xc functional in surface relaxations was unclear and worth 
    exploring in more detail. 
    
    In spite of the theoretical challenges to model and explain these metallic surface
    properties, density functional theory \cite{KOHN-DFT2, SCAN,SR16,norskov2011density,METAGGA} has proven to be one
    of the leading electronic structure theory methods to understand characteristics 
    of metal surfaces. Lang and Kohn \cite{KOHN-SURFACE, KOHN-WORK, KOHN-IMAGE} and 
    Skriver and Rosengaard \cite{SKRIVER-1} reported
    the surface energies and work functions of close-packed metal surfaces 
    from across the period table using Green's function
    techniques based on linear muffin-tin-orbitals  within the tight-binding 
    and atomic-sphere approximations. In another work, Perdew et al. used the 
    stabilized jellium and liquid drop model (SJM-LDM) \cite{perdew1991liquid} to understand the dependency
    of surface energies and work functions of simple metals on the bulk electron density
    as well as the atomic configuration of the exposed crystal face. Developing functionals that are
    accurate for surfaces has been an active 
    area of recent activity \cite{langreth1975exchange,perdew2008restoring,staroverov2004tests}.
    
    Although previous works \cite{LP77,pitarke2003metal,JelliumSurfEng-1} gave very reasonable descriptions of fundamental
    metallic surfaces, we must consider the limitations of the local and semilocal
    xc density functionals for metal surfaces \cite{kresse-shortcoming}. Wang et al.~\cite{Wang2014216} calculated surface
    energies and work functions of six close-packed fcc and bcc metal surfaces using
    LDA and PBE. Their study confirms the face-dependence of the surface energy
    and work function. Singh-Miller and Marzari \cite{Singh-Miller}  used PBE to study
    surface relaxations, surface energies, and work functions of the
    low-index metallic surfaces of Al, Pt, Pd, Au and Ti.
    Ref.~\onlinecite{Singh-Miller} found that LDA qualitatively agrees with the experimental 
    surface energies, but neither LDA or PBE can be considered as a default choice 
    for quantitative comparison with experimental values for surface properties. 
    Following what they have suggested, we will demonstrate that higher rungs of Jacob's ladder in 
    DFT \cite{MRS-JOHN}, such as meta-GGA's or the Random Phase Approximation \cite{langreth1975exchange}, 
    must be used to accurately study surface properties. 
    
    In this work we investigated the surface energies, work functions and interlayer
    relaxations of the low-index clean metallic surfaces of Al, Pt, Pd, Cu, Ag, Au, 
    Rh and Ru. Here we focus on three main crystallographic faces, (111), (110), and
    (100), to explore the face-dependence of the surface properties \cite{SD16}. Furthermore,
    we have explored the xc-functional dependence to demonstrate the improvements non-empirical meta-GGAs 
    can achieve compared to GGAs. We utilized the following approximations: LDA \cite{KOHN-DFT2},
    the PBE generalized gradient approximation and its modification for 
    solids, PBEsol \cite{perdew2008restoring,PBEsol}, and the newly constructed meta-GGAs SCAN and SCAN+rVV10.
    Pt (111) was used as a test case to explore the convergence of the surface  
    energy and work function with respect to slab thickness, kinetic energy cutoff,
    and k-points, see   Appendix~\ref{appendix:computational-details}. 

\section{Theory}
    \subsection{\label{sec:level2}Surface energy}
    The surface energy, $\sigma$, can be defined as the energy per atom, or per area, required
    to split an infinite crystal into two semi-infinite crystals with two equivalent 
    surfaces \cite{zanguillphysics},
    \begin{equation}
    \sigma=\frac{1}{2A}[ {E_{Slab}}-{\frac{{N_{Slab}}}{{N_{Bulk}}}}E_{Bulk}],
    \label{eq:surface-energy}
    \end{equation}
    where  $E_{Slab}$ is the total energy of the slab, $N$ is the total number of atoms in the slab,
    $E_{Bulk}$ is the total energy of the bulk, and $A$ is the surface area of the slab. The factor 
    of $1/2$ in the above equation comes from fact that each slab is bounded by two 
    symmetric surfaces.
    DaSilva et. al \cite{DaSilva} showed that a dense mesh can be used to avoid numerical instabilities 
    in Eq.~\eqref{eq:surface-energy} coming from using different numbers of atoms in the slab 
    and bulk calculations \cite{boettger}. The linear fit method is one way to find converged values of the surface energies from slab calculations \cite{boettger1998extracting}.
    Previously, Fiorentinni et. al ~\cite{methfessel}  applied this method for Pt (100) surface. We have applied this method to obtain converged values of the surface energies using 
    equivalent
    cutoff energies and dense k-meshes for bulk and surface calculations. One can write Eq.~\eqref{eq:surface-energy-2}
    for the large $N$ limit of the layer thickness as
    \begin{equation}
    E_{Slab} \approx N{E_{Bulk}}+2{\sigma}A \, ,
    \label{eq:surface-energy-2}
    \end{equation}
    so that the surface energy can be extracted from an extrapolation of the slab energy with
    respect to the layer thickness.

    \subsection{\label{sec:level2}Work function}
    The work function for metallic systems can be determined computationally taking
    the difference between the vacuum potential and the Fermi energy \cite{KOHN-WORK}:
    \begin{equation}
    \phi = V_{vacuum} - E_{Fermi} .
    \label{eq:work-function}
    \end{equation}
    The reported anisotropy of the work function for
    different surfaces implies it can depend on the particular face due to edge effects  \cite{work-anisotropy}.
    For different surfaces there are different densities of electrons at the edges
    of the surface and that causes different surface dipole barriers $D$,
    which is explicitly related to $V_{vacuum}$.
    Surface dependent values of $D$ can yield different values for the work function
    since the Fermi energy is solely a bulk property.
 
    \subsection{\label{sec:level2}Surface relaxations:}
    Surface relaxations arise due to the minimization of the energy at the surface
    and can be computed using the simple formula:
    $d_{ij}\% = \frac{{d_{i}}-{d_{j}}}{{d_{0}}}\times 100$,
    where ${d_{i}}$ and ${d_{j}}$ are the distance of $i^{th}$ and $j^{th}$
    layer from the top layer of the relaxed slab,
    and ${d_{0}}$ is the distance between the layers of the unrelaxed slab.

    \section{Results \& discussions:}
    
    \subsection{Surface energy}
    \label{sub:surface-energy}

        \begin{table*}[ht]
            \begin{ruledtabular}
            \begin{tabular}{cccccccc}
                \toprule
                \textbf{Metals}&\textbf{LDA}&\textbf{PBE}&\textbf{PBEsol}&\textbf{SCAN} &
                                \textbf{SCAN+rVV10} & \textbf{RPA}\footnotemark[3] & \textbf{Experimental}  \\   
                \midrule
                Al &  1.08 & 0.89 & 1.06 &    1.03 & 1.16   & 1.07  & 1.14\footnotemark[1]   \\ [0.2cm]       
                Cu &  1.98 & 1.48 & 1.74 & 1.68 & 1.89    & 2.04  & 1.83\footnotemark[2] \\ [0.2cm]         
                Ru &  3.19 & 2.48 & 2.89 & 2.77 & 2.99    & 3.45  & 2.99\footnotemark[1] \\ [0.2cm]         
                Rh &  2.86 & 2.47 & 2.71 & 2.6 & 2.81     & 3.10  & 2.7\footnotemark[2] \\ [0.2cm]          
                Pd &  2.19 & 1.59 & 1.90 & 1.8 &  2.04    & 2.32  & 2.00\footnotemark[1]   \\ [0.2cm]       
                Ag &  1.2 & 0.84 & 1.08 & 1.03 & 1.22     & 1.42  & 1.25\footnotemark[2]  \\ [0.2cm]        
                Pt &  2.26 & 1.79 & 2.12 & 1.92 & 2.15    & 2.7   & 2.49\footnotemark[1]     \\ [0.2cm]     
                Au &  1.41 & 0.87 & 1.16 & 1.06 & 1.29    & 1.41  & 1.5\footnotemark[1]  
            \end{tabular}
            \end{ruledtabular}
                \footnotetext[1]{Experimental results from Ref.~\onlinecite{al111surfexp}.}
                \footnotetext[2]{Experimental results from Ref.~\onlinecite{wangsurf111}.}
                \footnotetext[3]{\textsc{gpaw} results from Ref.~\onlinecite{BSR_surface}.}
            \caption{
                Mean surface energy of (111)[$\sigma_{111}$],(110)[$\sigma_{110}$] 
                and (100) [$\sigma_{100}$] surfaces of different metals in J/m$^{2}$.
            }
            \label{tab:avg-surf} 
            \end{table*}
    
    \begin{figure*}[htp]
        \includegraphics[width=\textwidth]{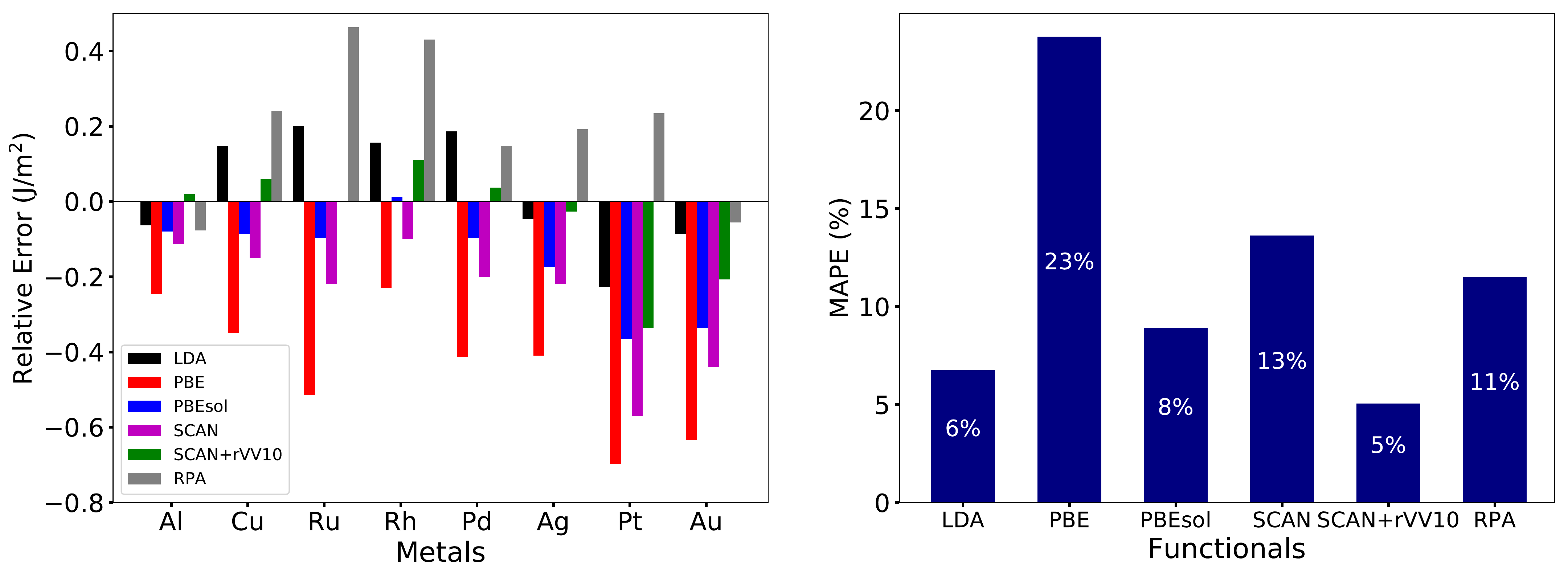}
        \caption{ 
            The relative error (RE) of the average surface energy (left) of the 
            (111), (110), and (100) surfaces compared to experiment \cite{al111surfexp,wangsurf111}. 
            The mean absolute percentage error (MAPE) of the surface energies (right) for each functional.
        }
        \label{fig:error-surf}    
    \end{figure*}    

    \begin{figure*}[htp]
        \includegraphics[width=1.0\textwidth]{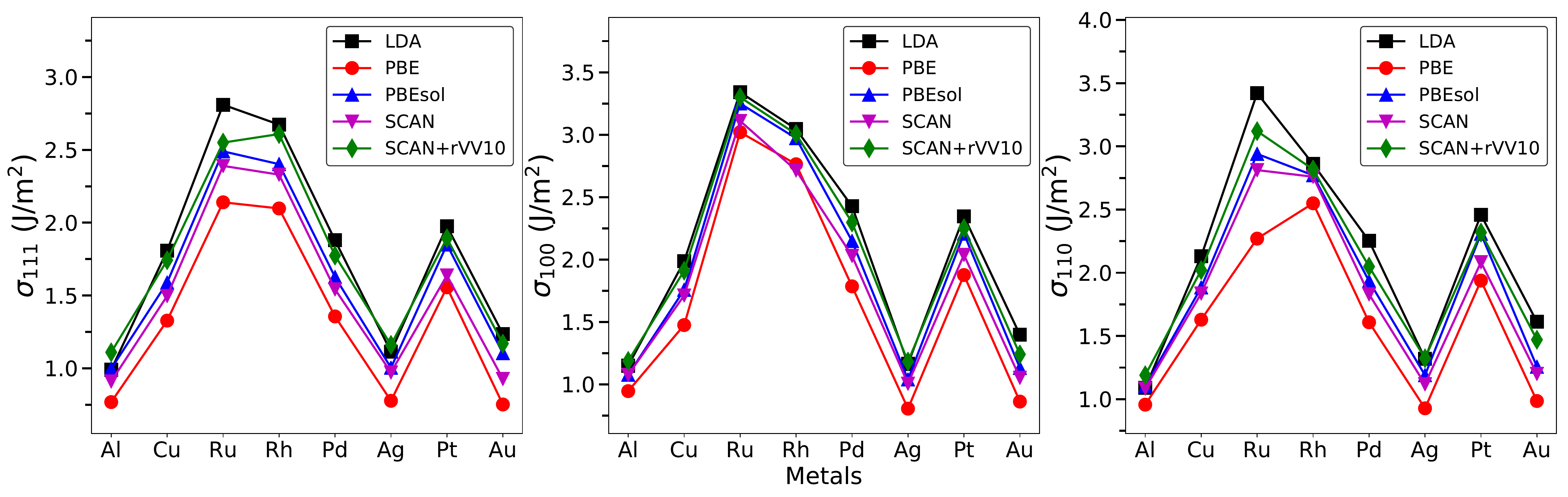}
        \caption{$\sigma_{111}$ (left), $\sigma_{110}$ (middle),  and $\sigma_{100}$ (right) 
            for the selected metals in this work.
        }
        \label{fig:face-surf}
    \end{figure*}

    The relevant surface energies measured experimentally \cite{al111surfexp}  are usually ``average''
    surface energies of all the possible surfaces present in the sample, hence experimentally
    measured surface energies should be compared with the average of the surface
    energies for (111), (110), and (100) surfaces \cite{Waele2016}. Here we also use the average
    surface energies to compare with the experimentally measured values, but from a
    different perspective. LDA is known to yield accurate surface energies for jellium,
    within the uncertainty of the latest QMC values \cite{QMC-jellium}, and an equally-weighted average over
    the three lowest-index faces from LDA reproduces the experimental surface energies
    to within their uncertainties. Many people have justified the accuracy 
    of LDA through an excellent error cancellation between its exchange and correlation
    contributions. Usually the LDA exchange energy contribution to the surface energy is an overestimate,
    while the correlation contribution is a significant underestimate, 
    and their combination results in an accurate prediction.
    
    In Table~\ref{tab:avg-surf} we report the mean surface energies calculated using different density
    functionals, including results from the random phase approximation (RPA) \cite{BSR_surface}. 
    Figure~\ref{fig:error-surf}(left) shows the relative error (in J/m$^2$) 
    of the computed values of the average surface energies compared to the best
    available experimental results for each metal. The consistent performance of SCAN+rVV10
    can be seen in all cases whereas PBE and SCAN both perform poorly. RPA results 
    are overall in good agreement with the experimental results, however the computational
    cost is higher and the improvement only marginal. One can argue that SCAN+rVV10 
    is the ``best'' candidate for predicting metallic surface energies with its moderate
    computational cost and high accuracy. 
    
    The relative errors and mean absolute percentage errors for the computed average surface energies are shown in 
    Figure \ref{fig:error-surf}. The relative errors are calculated with respect to the average experimental value 
    from all the three crystallographic surfaces. 
    Our results are in agreement within an acceptable margin compared to those previously 
    reported in the literature \cite{vitos1998surface,Singh-Miller,Waele2016,Wang2014216}.
    For Al, the lefthand plot illustrates the accuracy of all methods for simple metals that are close to the
    jellium limit. 
    Table \ref{tab:face-surf} demonstrates that there is an overall systematic improvement 
    from PBE to SCAN to SCAN+rVV10 in the Al surface energy due to the step-wise incorporation of 
    intermediate-range dispersion in SCAN and long-range dispersion in rVV10.
    The long-range contributions from rVV10 in Al account for 12\% of the total surface energy,
    and foreshadows the importance of including this contribution for the $d$ metals. 
    However, we find that the dispersion contribution from SCAN+rVV10 to the 
    total surface energy can be as large as 18\%. 
    Transition metal surfaces are more challenging due to their localized $d$ orbitals which cause inhomogeneities
    in the electron density at the surface.
    These inhomogeneities lead to a wider spread in the results from the different functionals.
    PBE yields the largest errors for the transition metal surface energies 
    due to its parametrization for slowly varying bulk densities.
    PBEsol was instead fit to jeliium surface exchange-correlation energies and yields a significant improvement
    compared to PBE.
    To improve the results further, vdW interactions need to be incorporated. 
    SCAN was constructed to interpolate the xc enhancement factor between covalent and metallic bonding limits
    in order to deliver an improved description of intermediate-range vdW interactions. 
    Consequently, SCAN is more accurate than PBE but not PBEsol, since no information 
    about surfaces was used in SCAN's construction. Still, even without parameterizing to jellium surfaces,
    SCAN and PBEsol are similar and exhibit analogous trends for all of the metals.
    With the addition of long-range vdW from rVV10, SCAN+rVV10
    surpasses the accuracy of PBEsol, indicating that the total vdW contribution to the surface energy
    is more important than previously recognized since SCAN alone is unable to outperform PBEsol.
    
    Although LDA does not explicitly include vdW interactions, we infer portions of the long-range part are somehow captured
    through error cancellation in the exchange and correlation contributions.
    RPA, which also includes vdW interactions, tends to overestimate the surface energies. 
    This is somewhat expected based upon the results for 
    jellium slabs \cite{PE98,KP99} where the xc contributions to the surface energy 
    show similar relative trends between LDA and RPA as a function of the Seitz radius. 

    The righthand plot in Figure~\ref{fig:error-surf} shows the mean absolute percentage errors (MAPE).
    SCAN+rVV10 is clearly the best semilocal density functional, though LDA is a close second.
    Incorporation of vdW interactions is important for dealing with the interactions of clean metallic surfaces and
    their surroundings, 
    however, and SCAN+rVV10 can be expected to perform more systematically than LDA for a broader set of properties.
    Though RPA provides a good additional benchmark when experimental data is scarce, its higher computational cost
    limits its applicability for general surface problems and reinforces the utility of a functional such as SCAN+rVV10,
    which is accurate, efficient, and naturally incorporates dispersion.
    
    Table~\ref{tab:face-surf} and Figure~\ref{fig:face-surf} illustrate the detailed performance of each method
    for each crystallographic face.
    SCAN+rVV10 frequently overlaps with LDA, 
    while the systematic underestimation of the surface energies by PBE is easy to see. 
    We find excellent agreement of our PBEsol results with that of Sun et al.~\cite{Jianwei-CO-PRB-2011},
    and that our LDA and PBE values and trends are in good agreement 
    with others recently reported \cite{Singh-Miller, Waele2016,Wang2014216}.
    The general trend of $\sigma_{111} < \sigma_{100} < \sigma_{110}$ can be seen from Fig.~\ref{fig:face-surf}
    for Al, Cu, Ag, Pt and Au respectively. However, this trend seems to be broken for Ru, Rh and Pd.

      \begin{table*}
    	\hfill{}
    	\begin{ruledtabular}
    		\begin{tabular}{ccccccccc}
    			\multirow{2}{*}{\textbf{Metals}}&\multirow{2}{*}{\textbf{Surface}}&\multirow{2}{*}{\textbf{LDA}}
    			&\multirow{2}{*}{\textbf{PBE}}&\multirow{2}{*}{\textbf{PBEsol}}
    			&\multirow{2}{*}{\textbf{SCAN}} &\textbf{SCAN} &\textbf{LDA} & \textbf{GGA} \\
    			& & & & & &\textbf{+rVV10} & (Other works) & (Other works) \\
    			\hline 
    			Al & 111 & 0.99 &0.77 & 0.99 &    0.91 & 1.11 & $0.91^{h}$ & $0.67^{g}$ \\ [0.2cm]
    			&  110 & 1.09 & 0.96 & 1.11 & 1.09 &1.19 & & $ 0.93^{g}$,$1.27^{b}$ \\ [0.2cm]
    			&  100 & 1.15 & 0.95 & 1.08 & 1.08 & 1.18 & &  $0.86^{g}$,$1.35^{b}$ \\[0.4cm]
    			
    			Cu & 111 & 1.81 & 1.33& 1.59 & 1.49 & 1.74 & &  \\ [0.2cm]
    			&  110 & 2.13 & 1.63 & 1.88 & 1.84  & 2.02 & $2.31^{d}$ & \\ [0.2cm]
    			&  100 & 1.99 & 1.48 & 1.76 & 1.71 & 1.91 & & $2.15^{e}$  \\[0.4cm]
    			
    			Ru & 111 & 2.81 & 2.14 & 2.49 & 2.39 & 2.55 & &  \\ [0.2cm]
    			&  110 & 3.42 & 2.27 & 2.94 & 2.81 & 3.12 & $3.45^{f}$ & \\ [0.2cm]
    			&  100 & 3.34 & 3.02 & 3.25 & 3.11 & 3.3  &  $3.52^{f}$ &   \\[0.4cm]
    			
    			Rh & 111 & 2.67 & 2.09 & 2.40 & 2.33 & 2.61  & &  \\ [0.2cm]
    			&  110 & 2.86 & 2.55 & 2.77 & 2.76 & 2.82 & $2.88^{f}$  &  \\ [0.2cm]
    			&  100 & 3.04 & 2.77 & 2.97 & 2.71 & 3.00 & $3.52^{f}$ &    \\[0.4cm]
    			
    			Pd & 111 & 1.88 & 1.36 & 1.63 & 1.54 &  1.77 & $1.87^{h}$ & $1.31^{g}$ \\ [0.2cm]
    			&  110 & 2.25 & 1.61 & 1.93 & 1.83 & 2.05 & $1.97^{f}$ & $1.55^{g}$ \\ [0.2cm]
    			&  100 & 2.43 & 1.79 & 2.15 & 2.03 & 2.29 & & $1.49^{g}$,$2.15^{e}$  \\[0.4cm]
    			
    			Ag & 111 & 1.13 & 0.78 & 1.00 & 0.97 &1.16 & & \\ [0.2cm]
    			&  110 & 1.32  & 0.93 & 1.19 & 1.12 & 1.33 & $1.26^{f} $ &  \\ [0.2cm]
    			&  100 & 1.16 & 0.81 & 1.04 & 1.00 & 1.18 & & $1.2^{b}$  \\[0.4cm]
    			
    			Pt & 111 & 1.98 & 1.56 & 1.85 & 1.64 & 1.89 & $2.23^{h}$ & $1.49^{g}$  \\ [0.2cm]
    			&  110 & 2.46 & 1.94 & 2.31 & 2.08 & 2.32 & & $1.85^{g}$,$2.49^{e}$   \\ [0.2cm]
    			&  100 & 2.35 & 1.88 & 2.21 & 2.04 & 2.25 & & $1.81^{g}$,$2.47^{e}$  \\[0.4cm]
    			
    			Au & 111 & 1.24 & 0.75 & 1.1 & 0.93 & 1.17  & $1.04^{g}$ & $0.74^{g}$  \\ [0.2cm]
    			&  110 & 1.61 & 0.99 & 1.26 & 1.2  & 1.47 & $1.55^{i}$ & $0.9^{g}$,$1.7^{b}$   \\ [0.2cm]
    			&  100 & 1.39 & 0.86 & 1.13 & 1.05 & 1.24 &  $1.39^{k}$ & $0.85^{g}$,$1.36^{e}$   
    		\end{tabular}
    	\end{ruledtabular}
    	\hfill{}
    	\caption{Surface energies (J/m$^{2}$) of the (111), (110), and (100) surfaces for the selected metals.}
    	\label{tab:face-surf} 
    	\begin{footnotesize}
    		a Experiment ; Ref.~\onlinecite{al111surfexp} \\
    		b FCD-GGA ; Ref.~\onlinecite{vitos1998surface} \\
    		c Experiment ; Ref.~\onlinecite{wangsurf111} \\
    		d Green function LMTO ; Ref.~\onlinecite{SKRIVER-1} \\
    		e PBE-GGA ; Ref.~\onlinecite{Wang2014216}  \\
    		f Full potential LMTO ; Ref.~\onlinecite{trendsmethfessel} \\
    		g PBE calculation ; Ref.~\onlinecite{Singh-Miller}\\
    		h All-electron LDA ; Ref.~\onlinecite{DaSilva}\\
    		i PWPP-LDA ; Ref.~\onlinecite{Alavi}\\
    		j PWPP-LDA ; Ref.~\onlinecite{Yu-Scheff}\\
    	\end{footnotesize}
    \end{table*}

 \subsection{Work function}
 \label{sub:work-function}
    
\begin{figure*}[htp]
    \includegraphics[width=1.0\textwidth]{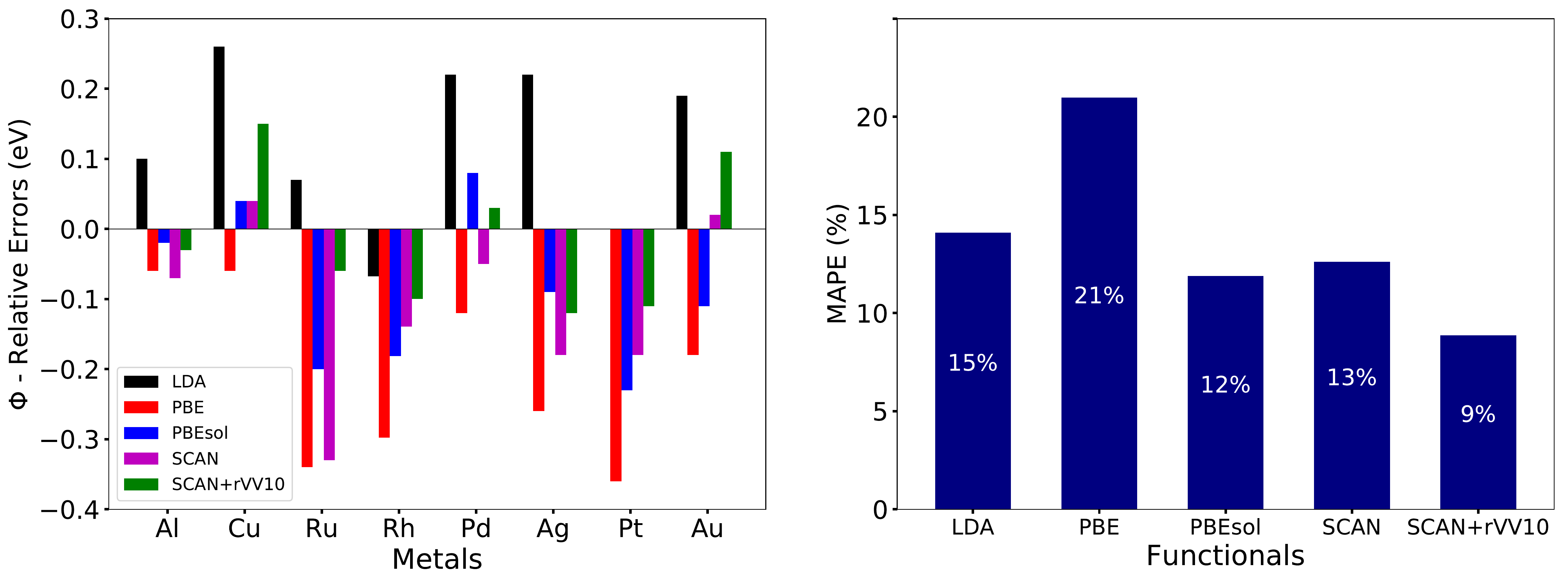}
  	\caption{
  		The relative errors in the work functions (top) predicted by 
  		each functional for the (111) surfaces. Mean absolute percentage
  		errors (MAPE) of the calculated work functions (bottom)  for the same systems.
  	}
  	\label{fig:111-work}
  \end{figure*}

\begin{figure*}[htp]
  	\includegraphics[width=1.0\textwidth]{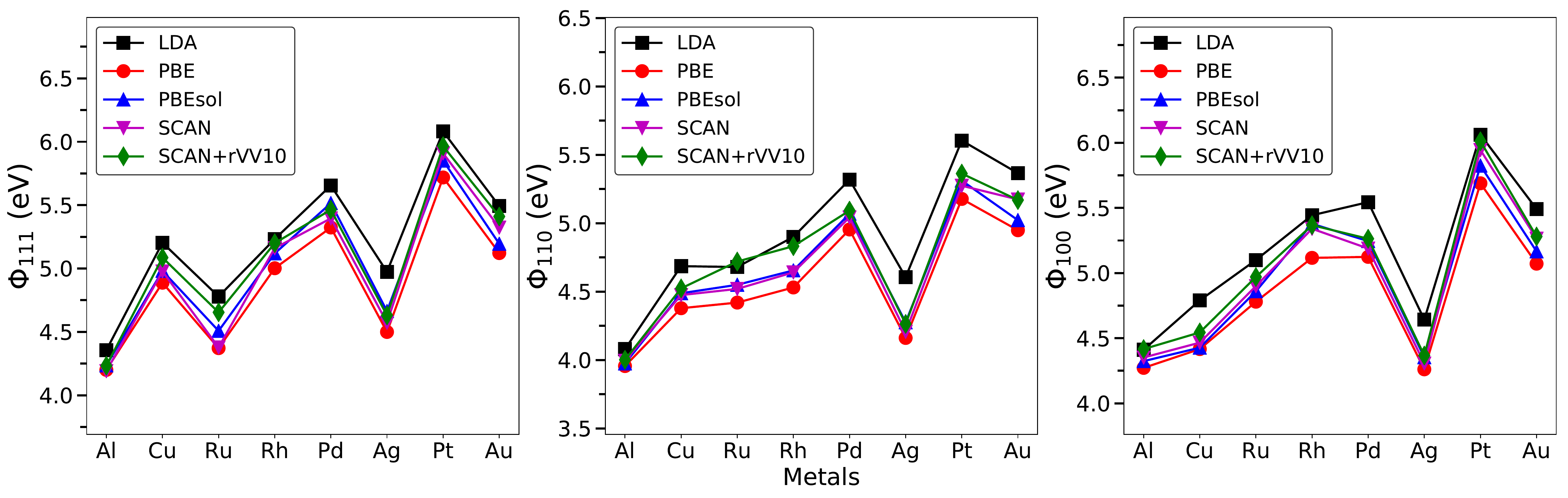}
  	\caption{
  		$\phi_{111}$ (left), $\phi_{110}$ (middle), and $\phi_{100}$ (right) for the surfaces studied in this work.
  	}
  	\label{fig:face-work}
  \end{figure*}

  
    The relative errors and MAPE with respect to experiment
    for the work functions of the (111) surfaces are plotted in Figure~\ref{fig:111-work}.
    Tabulated values of the work function for each face can be found in Table~\ref{tab:face-work}, 
    and are plotted in Fig.~\ref{fig:face-work}.
    Since we could not find experimental references for Ru (110) and (100) work functions,
    we instead focus on the (111) surface for simplicity.
    The performance trends for (111) generally hold for the other crystallographic faces as well.
    Our results for LDA and PBE are generally within $\approx0.15$ eV of those reported in the literature \cite{AlWFLDA,Singh-Miller}.
    For Al, LDA overestimates the work function for the (111) surface by 0.1 eV, but is dead on experiment for the other two faces.
    PBE and SCAN perform similarly for Al, but show larger deviations from one another for the $d$-block metals.
    PBEsol and SCAN+rVV10 yield the smallest errors for Al. The effect of geometric relaxation on the work function was
    not explored, but is likely negligible.

    Figure~\ref{fig:111-work} also shows the relative errors in the calculated values of the work function for the 
    transition metals. 
    These systems have entirely or partly filled $d$-orbitals which are localized on the atoms. 
    	
    Hybridization between the $d$ and $s$ orbitals varies with the crystallographic orientation resulting in changes
    in the surface-dipole and, consequently, the work function. The redistribution of the $d$ electrons in noble metals
    also impacts the surface energy due to changes in the bulk Fermi energy, 
    and these changes vary from one face to another \cite{Fall-2000-PRB}.

    The difference in work function values predicted by different functionals originates 
    from the different Fermi energies predicted by each functional. 
    The lower Fermi energy predicted by LDA results in an overestimated work function, 
    since the average electrostatic potentials calculated from different functionals differs by less than 0.2 eV from each other. 
    PBE and SCAN predict comparatively larger Fermi energies leading to underestimated work functions. 
    On the other hand, SCAN+rVV10 not only lowers the Fermi energy value but also increases the average electrostatic potential 
    compared to SCAN and thus gives a much better work function.
    From Fig.~\ref{fig:111-work} it is clear that PBE systematically underestimates the (111) work functions and its accuracy
    is erratic. 
    The underestimation of the (111) work function by PBE persists for the other faces as well.
    For Al and Cu PBE is accurate but shows much larger errors for the other metals. For Al and Ru, PBE and SCAN
    are fortuitously close though not for any particular physical reason, and in general SCAN improves upon PBE through
    its incorporation of vdW contributions to the surface potentials.  
    Though PBEsol and SCAN incorporate different physical limits in their construction, their overall performance for work
    functions is quite similar, and typically the errors from these functionals are within the experimental uncertainties.
    They are also outperforming LDA for the work functions, which was not the case for the surface energies above. 
    
    The inclusion of interemdiate-range vdW interactions
    is not enough, however, as the long-range contributions can still raise the work function by
    an appreciable amount. The (111) surface of Ru
    is one such case where the addition of rVV10 to SCAN increases the work function by nearly 0.3 eV, significantly reducing
    the error compared to experiment. Incorporating the long-range dispersion amounts to between 3 and 6\% of the total work function
    for the (111) surfaces, underscoring the importance of its inclusion.
    Though LDA and SCAN+rVV10 were of similar quality for the surface energies, SCAN+rVV10 clearly takes the top spot for computing
    accurate work functions of the (111) surfaces. 
    We note that the trend $ \phi_{110} < \phi_{100} < \phi_{111} $ predicted 
    by Smoluchowski \cite{work-anisotropy} is not observed for Ru, Rh and Pt, but is observed for the other metals.
    
    

    The photoelectrons ejected from the metal experience an image potential, which in DFT is influenced by
    the behavior of the exchange-correlation potential (functional). The attractive dispersion interaction lowers
    this potential, systematically increasing the work function.
    By incorporating a long-range contribution to the potential from rVV10, SCAN+rVV10 systematically and accurately
    predicts surface energies within experimental uncertainties.
    Addition of rVV10 to the GGAs would likely reduce their errors as well, provided the
    bare functional underestimates the experimental reference, but it would worsen the LDA results for all but Rh.
    The systematic behavior of SCAN for diversely bonded systems lends itself to correction by rVV10 achieving
    a well balanced performance both for surface and bulk \cite{SCAN,SR16} properties.

      Since, the surface energies are calculated using the relaxed slab model the effect of surface relaxations are already incorporated in the calculations. 

  \begin{table*}
  	\hfill{}
  	\begin{ruledtabular}
  		\begin{tabular}{cccccccccc}
  			\multirow{2}{*}{\textbf{Metals}}&\multirow{2}{*}{\textbf{Surface}}&\multirow{2}{*}{\textbf{LDA}}
  			& \multirow{2}{*}{\textbf{PBE}}&\multirow{2}{*}{\textbf{PBEsol}}&\multirow{2}{*}{\textbf{SCAN}} 
  			& \textbf{SCAN} &\textbf{LDA} & \textbf{GGA} & \multirow{2}{*}{\textbf{Expt.}} \\
  			\textbf{}&\textbf{}&\textbf{}&\textbf{}&\textbf{}&\textbf{} &\textbf{+rVV10} & (Other work) & (Other work)  & \textbf{} \\
  			\hline
  			Al& 111 & 4.36 & 4.2 & 4.24 & 4.19 & 4.23 & $4.25^{1}$ & $4.02^{2}$ &$4.26\pm0.03^{3}$,$4.32\pm 0.06^{4}$ \\ [0.2cm]
  			&  110 & 4.08 & 3.96 & 3.98 & 3.99 & 4.00 & $4.3^{1}$ & $4.3^{2}$ &$4.06 \pm 0.03^{5}$,$4.23 \pm 0.13^{4}$ \\ [0.2cm]
  			&  100 & 4.41 & 4.27 & 4.32 & 4.35 & 4.42 & $4.38^{1}$& $4.09^{2}$ &$4.41 \pm 0.03^{3}$,$4.32 \pm 0.06^{4}$ \\[0.4cm]
  			
  			Cu & 111 & 5.20 & 4.88 & 4.98 & 4.98 & 5.09 & & &$4.94^{6}$,$4.9 \pm 0.02^{4}$ \\ [0.2cm]
  			&  110 & 4.68 & 4.38 & 4.48 & 4.47 & 4.53 & & &$4.59^{7}$,$4.56 \pm 0.1 ^{4}$ \\ [0.2cm]
  			&  100 & 4.79 & 4.42 & 4.43 & 4.47 & 4.54 & & & $4.59 \pm 0.03^{8}$,$4.73 \pm 0.1 ^{4}$ \\[0.4cm]
  			
  			Ru & 111 & 4.78 & 4.37 & 4.51 & 4.38 &4.65 & $5.33^{9}$ &  & $4.71^{10}$ \\ [0.2cm]
  			&  110 & 4.68 & 4.42 & 4.55 & 4.52 &4.72 & $4.65^{9}$ & & \\ [0.2cm]
  			&  100 & 5.1 & 4.78 & 4.86 & 4.9 & 4.97 & $5.03^{9}$ &  & \\[0.4cm]
  			
  			Rh & 111 & 5.23 & 5.00 & 5.12 & 5.16 & 5.20 & & &$5.3 ^{11}$,$5.46 \pm 0.09^{4}$ \\ [0.2cm]
  			&  110 & 4.9 & 4.53 & 4.66 & 4.65 & 4.83 & $4.98^{9}$ & & $4.8 \pm 0.05 ^{12}$,$ 4.86 \pm 0.21 ^{4}$ \\ [0.2cm]
  			&  100 & 5.44 & 5.12 & 5.38 & 5.34 & 5.37 & $5.25^{9}$ & & $5.11 ^{13}$,$5.3 \pm, 0.15^{4}$ \\[0.4cm]
  			
  			Pd & 111 & 5.66 & 5.32 & 5.52 & 5.39 & 5.47 & $5.64^{14}$ & $5.25^{2}$ &$5.44 \pm 0.03^{15}$,$5.67 \pm 0.12^{4}$\\ [0.2cm]
  			&  110 & 5.32 & 4.95 & 5.07 & 5.04 & 5.09 &  & $4.87^{2}$ &$5.2^{16}$,$5.07 \pm 0.2^{4}$\\ [0.2cm]
  			&  100 & 5.54 & 5.12 & 5.25 & 5.19 & 5.26 & & $5.11^{2}$ &$5.3^{17}$,$5.48 \pm 0.23^{4}$ \\[0.4cm]
  			
  			Ag & 111 & 4.97 & 4.49 & 4.66 & 4.57 & 4.63 & & &$4.75 \pm 0.01^{18}$,$4.53 \pm 0.07^{4}$ \\ [0.2cm]
  			&  110 & 4.61 & 4.16 & 4.28 & 4.21 & 4.26 & & &$4.25 \pm 0.03^{19}$,$4.1 \pm 0.15^{4}$ \\ [0.2cm]
  			&  100 & 4.64 & 4.26 & 4.35 & 4.3 & 4.37 & & &$4.42 \pm 0.02^{20}$,$4.36 \pm 0.05^{4}$ \\[0.4cm]
  			
  			Pt & 111 & 6.08 & 5.72 & 5.85 & 5.90 & 5.97 & $6.06^{14}$ & $5.69^{2}$ &$6.08 \pm 0.15^{22}$,$5.91 \pm 0.08^{4}$ \\ [0.2cm]
  			&  110 & 5.6 & 5.18 & 5.31 & 5.27 & 5.36 & $5.52^{21}$ & $5.26^{2}$ &$5.4^{23}$,$5.53 \pm 0.13^{4}$ \\ [0.2cm]
  			&  100 & 6.06 & 5.69 & 5.82 & 5.94 & 6.01 &  & $5.66^{2}$ &$5.9^{24}$ ,$5.75 \pm 0.13^{4}$ \\[0.4cm]
  			
  			Au & 111 & 5.49 & 5.12 & 5.19 & 5.32 & 5.41 & $5.63^{25}$ & $5.15^{2}$ &$5.3-5.6^{26}$,$5.33 \pm 0.06^{4}$ \\ [0.2cm]
  			&  110 & 5.36 & 4.94 & 5.02 & 5.17 & 5.3 & $5.41^{25}$ & $5.04^{2}$ &$5.2^{27}$,$5.16 \pm 0.22^{4}$ \\ [0.2cm]
  			&  100 & 5.49 & 5.07 & 5.17 & 5.26 & 5.28 & $5.53^{25}$ & $5.1^{2}$ &$5.22 \pm 0.04^{27}$,$5.22 \pm 0.31^{4}$ 
  		\end{tabular}
  	\end{ruledtabular}
  	\hfill{}
  	\caption{Work function of (111) [$\phi_{111}$], (110) [$\phi_{110}$] and (100) [$\phi_{100}$]surfaces of different metals.}
  	\label{tab:face-work}
  	\begin{footnotesize}
  		
  		1 PWPPW-LDA  ; Ref.~\onlinecite{Fall-1998-PRB} \\
  		2 PBE  ; Ref.~\onlinecite{Singh-Miller} \\
  		3 Experiment  ; Ref.~\onlinecite{Al111100WF} \\
  		4 Experiment  ; Ref.~\onlinecite{WFALL} \\
  		5 Experiment  ; Ref.~\onlinecite{Al110WF} \\
  		6 Experiment  ; Ref.~\onlinecite{Cu111WF} \\
  		7 Experiment  ; Ref.~\onlinecite{Cu110WF} \\
  		8 Experiment  ; Ref.~\onlinecite{Cu100WF} \\
  		9 Full potential LMTO  ; Ref.~\onlinecite{trendsmethfessel} \\
  		10 Experiment  ; Ref.~\onlinecite{Cuworkexp} \\
  		11 Experiment   ; Ref.~\onlinecite{Rh111WF} \\
  		12 Experiment  ; Ref.~\onlinecite{Rh110WF} \\
  		13 Experiment  ; Ref.~\onlinecite{Rh100WF} \\    
  		14 FLAPW-LDA  ; Ref.~\onlinecite{DaSilva} \\
  		15 Experiment  ; Ref.~\onlinecite{Pd111WF} \\
  		16 Experiment  ; Ref.~\onlinecite{Pd110WF} \\
  		17 Experiment  ; Ref.~\onlinecite{Pd100WF} \\
  		18 Experiment  ; Ref.~\onlinecite{Ag111WF} \\
  		19 Experiment  ; Ref.~\onlinecite{Ag110WF} \\
  		20 Experiment  ; Ref.~\onlinecite{Ag100WF} \\
  		21 PWPP-LDA  ; Ref.~\onlinecite{Alavi} \\
  		22 Experiment  ; Ref.~\onlinecite{Pt111WF} \\
  		23 Experiment  ; Ref.~\onlinecite{Pt110WF} \\
  		24 Experiment  ; Ref.~\onlinecite{Pt100WF} \\
  		25 PWPP-LDA  ; Ref.~\onlinecite{Fall-2000-PRB} \\
  		26 Experiment  ; Ref.~\onlinecite{Au111WF} \\
  		27 Experiment  ; Ref.~\onlinecite{Au110100WF} \\
  		
  	\end{footnotesize}
  \end{table*}

      \subsection{Surface relaxation}
      \label{sub:relaxation}  
      
    At the surface, the presence of fewer neighboring ions can cause changes in the
    equilibrium positions of the ions due to the changes in the inter-atomic forces.
    Surface reconstructions have been measured for Ag, Pt and Au.~\cite{SurfReconstrAu,Au-110-d12-Expt,ReconsturctAUPT}
    In these kinds of experimental measurements surface reconstructions are possible since the
    top layers rearrange in order to reach a new equilibrium position, and hence can
    change the work function. Nevertheless, we can always compare the performance of
    one functional to another compared to the best available experimental values. 
  
    Tables~\ref{tab:111-relax}, \ref{tab:110-relax}, and \ref{tab:100-relax} 
    show the tabulated values of the percentage relaxation
    for the top 4 layers of the three surfaces.  Different exchange-correlation functionals 
    may predict different interlayer relaxations compared to 
    the experimental data \cite{alLEED,Pt111100workexp,SURFACE-EXPT2,Pt111d12,Aud12d23,Pdd12d23}. 
    It is important to note that for $d_{23}$\% and $d_{34}$\%
    we have found only a few experimental results to compare with.
    
    In most cases SCAN+rVV10 and SCAN predict reasonably accurate interlayer relaxations
    in comparison to the experimental results. However, the values for the (100) surfaces 
    of Cu, Pd and Ag calculated using the SCAN+rVV10 and SCAN are much lower than the experimental 
    results. Only PBEsol predicts a reliable value of $d_{12}$\% in these cases. 
    Notice that in many cases the percentage interlayer relaxation 
    values calculated by LDA are much larger than the experimental values. 
    Moreover, Tables~\ref{tab:111-relax}, \ref{tab:110-relax}, and \ref{tab:100-relax} 
    also show that the LDA and PBE results calculated  
    in this work are in agreement with Ref.~\onlinecite{Singh-Miller}.
    
    \section{Summary}
    We studied three important surface properties of metals using the local density 
    approximation, two generalized gradient approximations (PBE and PBEsol), and a 
    new meta-GGA (SCAN) with and without a van der Waals correction. The surface energy,
    work function and interlayer relaxation were calculated and compared with the best
    available experiential values. 
    The choice of exchange-correlation potential has a noticeable effect on the surface
    properties of metals especially on the surface energy \cite{SAHNIEXCSURFACE}. 
    The performance of SCAN is comparable to PBEsol, 
    which is a boon for meta-GGA development, since existing 
    GGA's struggle compared to LDA in predicting good surface energies and work functions \cite{Singh-Miller}. 
    van der Waals interactions are present at metallic surfaces and have   
    a non-negligible contribution to the surface energy and work function, but make  
    a smaller impact on the interlayer relaxation.  Ferri et al.~\cite{Car-PRL-2015}  also found that
    van der Waals corrections can improve surface properties. van der Waals interactions
    provide an attractive interaction between one half of a bulk solid and the other 
    when the halves are separated, so vdW interactions tend to increase the surface  
    energy. Typically vdW forces lower the energy of a neutral solid more than for 
    a charged system, so vdW interactions also tend to increase the work function. 
    LDA overestimates the intermediate-range vdW attraction but has no long-range component. 
    These two errors of LDA may cancel almost perfectly for surface properties. 
    Although it is well understood that LDA predicts an exponential decay rather than 
    the true asymptotic nature of the surface barrier potential, the discrepancy only 
    matters for cases where the surface states are extended into the vacuum, such as 
    in the case of photoemission and scanning tunneling microscopy. Surface energies and work functions are more 
    or less a property of the local electron density at the surface and hence LDA can
    give excellent results.

    PBE underestimates the intermediate range vdW and has no long-range vdW, so it underestimates 
    surface energies and work functions. PBEsol and SCAN have realistic intermediate-range
    vdW and no long-range vdW, so they are more accurate than PBE but not as good as 
    LDA for predicting surface properties. 
    The asymptotic long-range vdW interactions missing in semilocal functionals can make up to a 10\% difference
    in the surface energy or work function, which implies there is a limit to the accuracy of these methods.
    SCAN+rVV10 stands out in this regard as it is a balanced combination
    of the most advanced non-empirical semilocal functional to date and the flexible non-local vdW correction from rVV10. 
    In addition to delivering superior performance for layered materials \cite{SCAN+rVV10}, SCAN delivers high quality surface
    energies, work functions, and surface relaxations for metallic surfaces.
	SCAN+rVV10 includes realistic 
	intermediate- and long-range vdW interactions, so it tends to yield more systematic and accurate
	results than LDA, PBEsol, or SCAN. 
    Accurate measurements for these properties 
    are needed in order to validate the performance of new and existing density functionals. 
    Overall we find that SCAN is a systematic step up in accuracy from PBE and that
    adding rVV10 to SCAN yields a highly accurate method for diversely bonded systems.
 

    \begin{table}[htp]
        \begin{ruledtabular}
            \begin{tabular}{ccccccc}
                {\textbf{Surface} } & {\textbf{LDA}} & {\textbf{PBE}
                } & {\textbf{PBEsol}} & {\textbf{SCAN}} & {\textbf{SCAN+}} \\
                {\textbf{} } & {\textbf{}} & {\textbf{}
                } & {\textbf{}} & {\textbf{}} & {\textbf{rVV10}} \\
                111   & 0.174 & 0.629 &  0.356 & 0.463 & 0.238   \\
            \end{tabular}
            \caption{MAE (J/m$^{2}$) in calculated (111) surface energies compared to the experimental values.}
            \label{tab:mae-111-surf}
        \end{ruledtabular}
    \end{table}

\begin{table}[H]
    \begin{ruledtabular}
            \begin{tabular}{ccccccc}
                {\textbf{Surface} } & {\textbf{LDA}} & {\textbf{PBE}
                } & {\textbf{PBEsol}} & {\textbf{SCAN}} & {\textbf{SCAN+}} \\
                {\textbf{} } & {\textbf{}} & {\textbf{}
                } & {\textbf{}} & {\textbf{}} & {\textbf{rVV10}} \\
                111   & 0.141 & 0.209 &  0.119 & 0.126 & 0.089   \\
            \end{tabular}
            \caption{MAE (eV) in work function of (111) surfaces compared to the experimental results.}
            \label{tab:mae-111-work}
            \end{ruledtabular}
        \end{table}
    
    \section{Acknowledgment}
    A.P. would like to thank A. Ruzsinszky, H. Peng, Z. Yang, and C. Shahi
    for their help and suggestions.  This work is supported by NSF under DMR-1305135,
    CNS-09-5884, and by DOE under DE-SC0012575, DE-AC02-05CH11231.

    \begin{center}
        \begin{table*}
            \hfill{}
            \begin{ruledtabular}
            \begin{tabular}{ccccccccccccc}
                \multirow{2}{*}{\textbf{Metals}}&\multirow{2}{*}{\textbf{Surface}}&\multirow{2}{*}{\textbf{LDA}}
                &\multirow{2}{*}{\textbf{PBE}}&\multirow{2}{*}{\textbf{PBEsol}}&\multirow{2}{*}{\textbf{SCAN}} 
                &\textbf{SCAN} &\textbf{LDA} & \textbf{PBE} &\textbf{Expt.} \\
                & & & & &  &\textbf{+rVV10} & (Other work) & (Other work)  &   \\ [0.1cm]
                \hline
                Al& {$d_{12}$\%} & 1.64 & 1.46 & 1.55 & 1.81 & 1.89 & $+1.35^{g}$ & $+1.04^{f}$  &$1.7 \pm 0.3^{i}$ \\ [0.2cm]            
                &  {$d_{23}$\%} & -0.66 & -0.72 & -0.73 & -1.27 & -1.27 & $+0.54^{g}$ & $-0.54^{f}$ & $-0.5\pm0.7^{i}$\\ [0.2cm]        
                & {$d_{34}$\%} & 0.1 & 0.07 & 0.09 & 0.17 & 0.16& $+1.04^{g}$ & $+0.19^{f}$  &   \\[0.4cm]
                
                Cu& {$d_{12}$\%} & -0.44 & -0.34 & -0.39 & -0.39 & -0.51 &  &  &$-0.7 \pm 0.5^{b}$ \\ [0.2cm]
                &  {$d_{23}$\%} & -5.43 & 0.01 &  0.1 & -0.1  & 0.14 &  & &  \\ [0.2cm]
                &  {$d_{34}$\%} & -4.8 & -0.01 & -0.1 & 0.13  &  -0.09 &  & &\\[0.4cm]
                
                Ru & {$d_{12}$\%} & -16.73 & -18.41 & -19.84 & -19.82 &  -17.26 &  & & \\ [0.2cm]
                &  {$d_{23}$\%} & -15.35 & -9.58 & -11.39 & -12.06 & -15.8 &  & &\\ [0.2cm]
                &  {$d_{34}$\%} & -3.74 & -8.43 & 5.84 & 6.11 & -4.01 &   & & \\[0.4cm]
                
                Rh & {$d_{12}$\%} & -1.24 & -1.93 & -1.56& -1.53 & -1.37 &  & & \\ [0.2cm]
                &  {$d_{23}$\%} & -0.5 & -0.89 & -0.36 & -0.2 & -0.72 &  & & \\ [0.2cm]
                &  {$d_{34}$\%} & -1.11 & 1.05 & 1.07 & 1.16 & 1.29 &  & &\\[0.4cm]
                
                Pd & {$d_{12}$\%} & -0.45 & -0.45 & 0.55 & 0.91 & 1.07 & $-0.22^{g}$ & $+0.25^{f}$ & $+1.3 \pm 1.3^{c}$ \\ [0.2cm]
                &  {$d_{23}$\%} & -0.52 & -0.24 & -0.3 & -0.49 & -0.42 & $-0.53^{g}$ & $-0.34^{f}$ &  $-1.3 \pm 1.3^{c}$\\ [0.2cm]
                &  {$d_{34}$\%} & -0.48 & 0.15 & 0.11 & 0.2 & 0.14 & $-0.33^{g}$ & $+0.10^{f}$ &   \\[0.4cm]
                
                Ag & {$d_{12}$\%} & 0.15 & -0.15 & -0.07 & -0.43 & 0.24 & $-0.53^{h}$ & $-0.3^{h}$ & $-0.5\pm 0.3^{d}$ \\ [0.2cm]
                &  {$d_{23}$\%} & -0.11 & -0.3 & -0.07 & -0.12 & -0.27 & $-0.07^{h}$  & $-0.04^{h}$ & $-0.4\pm 0.4^{d}$\\ [0.2cm]
                & {$d_{34}$\%} & -0.14 & -0.15 & -0.8 & -0.24 & -0.49 & $0.22^{h}$ & $0.16^{h}$ & $0 \pm 0.4^{d}$\\[0.4cm]
                
                Pt & {$d_{12}$\%} & 1.05 & 0.89 & 0.79 & 2.48 & 2.67 & $0.88^{g}$ &$+0.85^{f}$ & $ +1.1 \pm 4.4^{e}$ \\ [0.2cm]
                & {$d_{23}$\%} & -0.32 & -0.71 & -0.65 & -0.39 &  -0.15  & $-0.22^{g}$ & $-0.56^{f}$ & \\ [0.2cm]
                & {$d_{34}$\%} & 0.14 & -0.04 & -0.03 & -0.62 &  -0.61 & $-0.17^{g}$ & $-0.15^{f}$ &  \\[0.4cm]
                
                Au & {$d_{12}$\%} & -0.42 & 0.99 & 0.76 & 1.09 & 1.5 & $0.8^{g}$ & $-0.04^{f}$ &  \\ [0.2cm]
                &  {$d_{23}$\%} & -0.58 & -0.75 & -0.65 & -0.78 & -0.82 & $-0.3^{g}$ & $-1.86^{f}$ &  \\ [0.2cm]
                &  {$d_{34}$\%} & -0.24 & -0.29 & -0.18 & -0.27 & -0.31 & & $-1.4^{f}$ & \\[0.4cm]
            \end{tabular}
            \end{ruledtabular}
            \hfill{}
            \caption{Inter-layer relaxations for the (111) surfaces of different metals.} 
            \label{tab:111-relax}
            \begin{footnotesize}
                a LEED ; Ref.~\onlinecite{alLEED} \\
                b LEED ; Ref.~\onlinecite{Cu111d12} \\
                c LEED ; Ref.~\onlinecite{AgLEED} \\ 
                d LEED ; Ref.~\onlinecite{Pdd12d23} \\
                e LEED ; Ref.~\onlinecite{Pt111d12} \\
                f PBE ; Ref.~\onlinecite{Singh-Miller} \\
                g FLAPW-LDA ; Ref.~\onlinecite{DaSilva} \\
                h DFT(LDA \& PBE) ; Ref.~\onlinecite{Ag_relaxation}\\
            \end{footnotesize}
        \end{table*}
    \end{center}

    \begin{center}
        \begin{table*}[htp]
            \hfill{}
            \begin{ruledtabular}
            \begin{tabular}{ccccccccccccc}
                \multirow{2}{*}{\textbf{Metals}}&\multirow{2}{*}{\textbf{Surface}}&\multirow{2}{*}{\textbf{LDA}}
                &\multirow{2}{*}{\textbf{PBE}}&\multirow{2}{*}{\textbf{PBEsol}}&\multirow{2}{*}{\textbf{SCAN}} 
                &\textbf{SCAN} &\textbf{LDA} & \textbf{PBE} &\textbf{Expt.} \\
                & & & & &  &\textbf{+rVV10} & (Other work) & (Other work)  &   \\ [0.1cm]
                \hline
                Al& {$d_{12}$\%} & -7.04 & -7.27 & -6.84 & -8.86 & -8.44 & $-6.9^{g}$ & $-5.59^{f}$  &$-8.5 \pm 1.0^{a}$ \\ [0.2cm]
                
                &  {$d_{23}$\%} & 5.28 & 4.1 & 3.99 & 6.05 & 3.88 & & $+2.2^{f}$ & $+5.5\pm 1.1^{a}$  \\ [0.2cm]
                & {$d_{34}$\%} & -1.02 & -0.86 & -0.91 & -1.12 & -0.79 & $+2.2\pm 1.3^{g}$ & $-1.29^{f}$ &  $+2.2 \pm 1.3 ^{a}$  \\[0.4cm]
                
                Cu& {$d_{12}$\%} & -9.9 & -9.98 & -10.07 & -11.75 & -11.17 &  & &$-10 \pm 2.5^{b}$ \\ [0.2cm]
                &  {$d_{23}$\%} & 5.26 & 4.81 & 5.00 & 5.82  & 5.73&  &  &  $0 \pm 2.5^{b}$  \\ [0.2cm]
                &  {$d_{34}$\%} & -2.91 & -1.18 & -0.99 & -3.89  &  -2.6 &  & &\\[0.4cm]
                
                Ru & {$d_{12}$\%} &  -18.44  & -19.8 & -14.7 & -15.75 & -17.2 & & & \\ [0.2cm]
                &  {$d_{23}$\%} &  -9.65 & -6.23 & -8.55 & -9.41 & -9.24 &  & &\\ [0.2cm]
                &  {$d_{34}$\%} & -2.01 & 0.93 & 1.64 & -0.88 & -1.85 &  & & \\[0.4cm]
                
                Rh & {$d_{12}$\%} & -14.2 & -10.54 & -8.97 & -11.33 & -11.25 &  & & \\ [0.2cm]
                &  {$d_{23}$\%} & 2.74 & 2.49 & 3.09 & 2.87 & 3.57 &  & & \\ [0.2cm]
                &  {$d_{34}$\%} & -1.45 & 1.41 & 1.91 & 3.26 & 3.89 &  & &\\[0.4cm]
                
                Pd & {$d_{12}$\%} & -6.88 & -5.38 & -8.88 & -9.5 & -7.6 & $-5.3^{g}$ & $-8.49^{f}$  & $-5.8 \pm 2.2^{c}$ \\ [0.2cm]
                &  {$d_{23}$\%} & 4.03 & 3.84 & 4.11 & 4.73 & 4.00 &  & $+3.47^{f}$ & $ +1.0 \pm 2.2 ^{c}$ \\ [0.2cm]
                &  {$d_{34}$\%} & -0.35 & -0.21 & -0.3 & -0.28 & -0.44 & &  $-0.19^{f}$ & \\[0.4cm]
                
                Ag & {$d_{12}$\%} & -7.71 & -6.87 & -8.84 & -8.81 & -7.38 & $-8.81^{j}$ & $-9.19^{j}$ & $-7.8\pm 2.5^{d}$ \\ [0.2cm]
                &  {$d_{23}$\%} & 4.71 & 3.69 & 4.45 & 3.99 & 4.16 & $3.59^{j}$  &  $4.1^{j}$ & \\ [0.2cm]
                & {$d_{34}$\%} & -1.07 & -0.97 & -1.23 & -0.86 & -0.42 &  $-1.11^{j}$ & $-1.5^{j}$ & \\[0.4cm]
                
                Pt & {$d_{12}$\%} & -16.47 & -17.15 & -16.1 & -24.5 & -23.3 & & $-15.03^{f}$  & $ -18.5 \pm 2.2^{e}$ \\ [0.2cm]
                & {$d_{23}$\%} & 8.96 & 10.08 & 8.85 & 14.37 &  14.55 &  & $+7.61^{f}$ & $-24.2 \pm 4.3^{e}$ \\ [0.2cm]
                & {$d_{34}$\%} & -1.82 & -1.91 & -1.82 & -2.53 &  -2.23 & & $-1.7^{f}$ &  \\[0.4cm]
                
                Au & {$d_{12}$\%} & -14.08 & -13.87 & -13.52 & -14.54 & -14.09 & $-9.8^{g}$ & $-12.94^{f}$ & $ -20.1 \pm 3.5^{i}$\\ [0.2cm]
                &  {$d_{23}$\%} & 9.01 & 9.24 & 8.71 & 10.11 & 10.14 & $-7.8^{g}$ & $+7.83^{f}$ &  $ -6.2 \pm 3.5 ^{i}$\\ [0.2cm]
                &  {$d_{34}$\%} & -4.00 & -3.29 & -3.58 & -4.17 & -3.68 & $-0.8^{g}$ & $-2.66^{f}$ & 
            \end{tabular}
            \end{ruledtabular}
            \hfill{}
            \caption{Inter-layer relaxations for the (110) surfaces of different metals.} 
            \label{tab:110-relax}
            \begin{footnotesize}
                a LEED ; Ref.~\onlinecite{Al-110-d12-d23-Expt} \\
                b LEED ; Ref.~\onlinecite{Cu-110-d12-Expt} \\
                c LEED ; Ref.~\onlinecite{Pd-110-d12-d23-Exp} \\
                d LEED ; Ref.~\onlinecite{Ag-110-d12-Expt} \\
                e LEED ; Ref.~\onlinecite{Pt111d12} \\
                f PBE ; Ref.~\onlinecite{Singh-Miller} \\
                g PWPP-LDA ; Ref.~\onlinecite{Al-110-d12-LDA} \\
                h LDA-SGF ; Ref.~\onlinecite{trendsmethfessel} \\
                i LEED ; Ref.~\onlinecite{Au-110-d12-Expt} \\
                j DFT(LDA \& PBE) ; Ref.~\onlinecite{Ag_relaxation}\\
            \end{footnotesize}
        \end{table*}
    \end{center}

    \begin{center}
        \begin{table*}[htp]
            \hfill{}
            \begin{ruledtabular}
            \begin{tabular}{ccccccccccccc}
                \multirow{2}{*}{\textbf{Metals}}&\multirow{2}{*}{\textbf{Surface}}&\multirow{2}{*}{\textbf{LDA}}
                &\multirow{2}{*}{\textbf{PBE}}&\multirow{2}{*}{\textbf{PBEsol}}&\multirow{2}{*}{\textbf{SCAN}} 
                &\textbf{SCAN} &\textbf{LDA} & \textbf{PBE} &\textbf{Expt.} \\
                & & & & &  &\textbf{+rVV10} & (Other work) & (Other work)  &   \\ [0.1cm]
                \hline
                Al& {$d_{12}$\%} & 1.22 & 1.18 & 0.9 & 1.08 & 1.26 & $+0.5^{g}$ & $+1.73^{g}$ & $1.8^{a}$  \\ [0.2cm]
                &  {$d_{23}$\%} & -0.44 & -0.65 & -0.38 & 0.0 & -0.06 & & $0.47^{g}$ & \\ [0.2cm]
                & {$d_{34}$\%} & -0.32 & -0.26 & -0.24 & -0.67 & -0.69 & & $-0.27^{g}$ &   \\[0.4cm]
                
                Cu& {$d_{12}$\%} & -2.88 & -2.48 & -2.18 & -3.98 & -3.19 & & &$-1.1 \pm 0.4 ^{b}$  \\ [0.2cm]
                &  {$d_{23}$\%} & -0.41 & 1.22 & 1.15 & -0.25  & -0.75 & & & $1.7 \pm 0.6 ^{b}$  \\ [0.2cm]
                &  {$d_{34}$\%} & 0.0 & 0.11 & 0.03 & -0.13  &  -0.03 & & & \\[0.4cm]
                
                Ru & {$d_{12}$\%} & -17.71 & -13.12 & -15.79 & -15.85 & -15.65 & & &    \\ [0.2cm]
                &  {$d_{23}$\%} & -2.24 & 0.4 & -1.69 & -2.76 & -2.52  & & &   \\ [0.2cm]
                &  {$d_{34}$\%} & 0.83 & 3.29 & 1.32 & 1.47 & 1.39 & & &   \\[0.4cm]
                
                Rh & {$d_{12}$\%} & -7.71 & -4.13 & -3.63 & -4.45 & -4.43 & & & \\ [0.2cm]
                &  {$d_{23}$\%} & -2.39 & 0.47 & 0.81 & 1.37 & 1.83 & & & \\ [0.2cm]
                &  {$d_{34}$\%} & -2.23 & 0.92 & 1.14 & 0.98 & 1.32 & & & \\[0.4cm]
                
                Pd & {$d_{12}$\%} & -0.69 & -1.13 & -1.17 & -0.94 & -0.9 & $-0.6^{i}$ & $-1.3^{g}$ & $3.0 \pm 1.5 ^{c}$   \\ [0.2cm]
                &  {$d_{23}$\%} & 0.28 & 0.24 & 0.43  & 0.32 & 0.21 & & $0.0^{g}$ & $+1.0 \pm 1.5^{c}$\\ [0.2cm]
                &  {$d_{34}$\%} & -0.54 & 0.26 & 0.23 & 0.49 & 0.59 & & $+0.35^{g}$ & \\[0.4cm]
                
                Ag & {$d_{12}$\%} & -1.13 & -1.74 & -1.41 & -1.78 & -1.16 & $-1.81^{k}$ & $-1.87^{k}$ & $0\pm 1.5^{d}$ \\ [0.2cm]
                &  {$d_{23}$\%} & 0.87 & 0.79 &  0.69 & 0.89 & 0.83 & $0.56^{k}$ & $0.51^{k}$ & \\ [0.2cm]
                & {$d_{34}$\%} & 0.17 & 0.2 & 0.16 & -0.06 & 0.28 & $0.42^{k}$ & $0.3^{k}$ & \\[0.4cm]
                
                Pt & {$d_{12}$\%} & -2.61 & -2.2 & -3.84 & -4.29 & -3.72& & $-2.37^{g}$ & $ +0.2 \pm 2.6^{e}$ \\ [0.2cm]
                & {$d_{23}$\%} & -0.04 & 0.03 & 0.38 & -0.96 &  -0.96 & & $-0.55^{g}$ & \\ [0.2cm]
                & {$d_{34}$\%} & -1.66 & -1.35 & -1.39 & -1.37 &  -0.77 & & $+0.29^{g}$ &  \\[0.4cm]
                
                Au & {$d_{12}$\%} & 0.88 & 0.52 & 0.6 & 0.54 & 0.53 & $-1.2^{g}$ & $-1.51^{g}$ & $ -20 \pm 3^{f}$\\ [0.2cm]
                &  {$d_{23}$\%} & -0.721 & -0.75 & -0.65 & -0.78 & -0.82 & $0.4^{g}$ & $0.33^{g}$ & $+2 \pm 3^{f}$\\ [0.2cm]
                &  {$d_{34}$\%} & 0.44 & 0.24 & 0.17 & 0.22 & 0.19 & & $0.24^{g}$ & 
            \end{tabular}
            \end{ruledtabular}
            \hfill{}
            \caption{Inter-layer relaxations for the (100) surface of different metals.} 
            \label{tab:100-relax}
            \begin{footnotesize}
                a LEED ; Ref.~\onlinecite{Al-100-d12-Expt} \\
                b LEED ; Ref.~\onlinecite{Cu-100-d12-Expt} \\
                c LEED ; Ref.~\onlinecite{Pd-100-d12-Expt} \\ 
                d LEED ; Ref.~\onlinecite{Ag-100-d12-Expt} \\
                e LEED ; Ref.~\onlinecite{Pt-100-d12-Expt} \\
                f Hex XRD ; Ref.~\onlinecite{Aud12d23} \\
                g PBE ; Ref.~\onlinecite{Singh-Miller} \\
                h PWPP-LDA ; Ref.~\onlinecite{Al-100-d12-LDA} \\
                i  LDA-SGF ; Ref.~\onlinecite{trendsmethfessel} \\
                j PWPP-LDA ; Ref.~\onlinecite{Yu-Scheff}\\
                k DFT(LDA \& PBE) ; Ref.~\onlinecite{Ag_relaxation}\\
            \end{footnotesize}
        \end{table*}
    \end{center}
    \section{Appendices}
    \label{appendix:}

    \subsection{Stabilized jellium model}
    The jellium model (JM) studied in Refs.~\onlinecite{KOHN-SURFACE, KOHN-WORK, Manninen-Jellium}
    is a simple model to study surface properties in 
    which the positive ionic charge is replaced by a uniform positive background truncated
    at a planar surface. Although the jellium model shows ``universality'' in predicting 
    the $r_s$ dependence of metallic surface properties, it's not perfect. It has the following
    defects:
    \begin{enumerate}
        \item Negative surface energy for $r_s \approx 2$,\cite{KOHN-SURFACE}
        \item Negative bulk modulus for $r_s \approx 6$.~\cite{Ashcroft-pseudo}
    \end{enumerate}

    These defects are corrected in the SJM using a ``structureless pseudopotential'' \cite{SJM-1}. 
    SJM treats the ``differential potential'' between the pseudopotential of the lattice 
    and the electrostatic potential of the uniform positive background  perturbatively, 
    adapting the idea that each bulk ion belongs to a 
    neutral Wigner-Seitz sphere of radius $r_{0}$ with $r_{0}={z^{1/3}}{r_{s}}$.
    
   \begin{table}[H]
       \begin{ruledtabular}
            \hfill{}
            \begin{tabular}{ccccccccc}
                \textbf{}&\textbf{surface}
                & \textbf{JM} & \textbf{SJM}  & \textbf{SJ-LDM}  & \textbf{Expt}\\
                Al & 111 &  $-.605^{a}$ & $0.953$ & $1.096^{b}$  &$1.14^{a}$  \\ [0.2cm] 
            \end{tabular}
            \hfill{}
            \caption{Surface energies (J/m$^{2}$) of (111)surfaces of Al from different jellium models. }
            \label{tb:surf_al_jellium}
            \end{ruledtabular}
        \end{table}

\begin{table}[H]
    \begin{ruledtabular}
           
            \begin{tabular}{ccccccccc}
                \textbf{}&\textbf{surface}
                & \textbf{JM} & \textbf{SJM}  & \textbf{SJ-LDM}  & \textbf{Expt}\\
                Al & 111 &  $3.74^{a}$ & 4.24 & $4.09^{b}$  & $4.3^{a}$  \\ [0.2cm] 
            \end{tabular}
            \caption{Tabulated values of work function (eV) of (111) surface of Al using different jellium models.}
            \label{tb:work_al_jellium}
            \begin{footnotesize}
                a Ref.~\onlinecite{JelliumSurfEng-1,JelliumSurfEng-2} \\
                b Ref.~\cite{Simpletheory-simplemetal} \\
            \end{footnotesize}
      \end{ruledtabular}
        \end{table}
    
    \subsection{Jellium surface energies from semilocal density functionals} 
    \label{appendix:jellium-surface-energy}
     Jellium surface energies for different $r_s$
     can depends on the semilocal exchange-correlation functional. In this section,
     we tabulated the calculated values of jellium surface energies for LDA, PBE, PBEsol
     and SCAN. We mention here that the rVV10 long-range correction to SCAN is less
     important for the jellium surface than a real surface. This preserves the fact
     that the jellium surface is an appropriate norm for SCAN itself.

    \begin{table}[H]
        \begin{ruledtabular}
            \hfill{}
            \begin{tabular}{cccccc}
                \textbf{$r_{s}$}&\textbf{Exact }&\textbf{LDA}&\textbf{PBE}&\textbf{PBEsol} & \textbf{SCAN}\\
                \textbf{}&\textbf{ ($erg/cm^{2}$)}&\textbf{}&\textbf{}&\textbf{} & \textbf{}\\
                2 & 2624 & 15.7 & -7.2 & 1.6 & 0.3\\ 
                3 & 526 & 27.2 & -11.6 & 2.7 & -7.0\\
                4 & 157 & 42.7 & -18.5 & 3.2 & -19.1\\
                6 & 22 & 15.7 & -46.4 & 4.1 & -71.4\\
            \end{tabular}
            \hfill{}
            \caption{
                Exchange energies ($[{{\sigma}_{EX}}-{{\sigma}_{Exact-EX}}]/{{\sigma}_{Exact-EX}}$) of 
                jellium surface for different values of $r_{s}$.
            }
            \label{tb:exchange_jellium}
            \end{ruledtabular}
        \end{table}

 \begin{table}[H]
     \begin{ruledtabular}
            \hfill{}
            \begin{tabular}{ccccc}
                \textbf{Error}&\textbf{LDA}&\textbf{PBE}&\textbf{PBEsol} & \textbf{SCAN}\\
                \textbf{}&\textbf{}&\textbf{}&\textbf{} & \textbf{}\\
                ME (Ha)      &   160.89 & -72.3 & 15.53 & -18.66\\
                MARE (Ha)    &   160.89 & -72.3 & 15.53 & 22.60\\
                MRE (\%) &   45.83 & -20.93 & 2.90 & -24.30\\
                MARE (\%) &   45.83 & -20.93 & 2.90 &  24.45 \\
                RMSD (\%) &   36.31 & 17.61 & 1.04 &  32.40\\
            \end{tabular}
            \hfill{}
            \caption{
                Error in Exchange energies of jellium surface for different values of $r_{s}$ 
                calculated for the values in Table~\ref{tb:exchange_jellium}.
            }
            \label{tb:error_exchange_jelllium}
            \end{ruledtabular}
        \end{table}

 \begin{table}[H]
     \begin{ruledtabular}
            \hfill{}
            \begin{tabular}{ccccccc}
                \textbf{$r_{s}$}&\textbf{TDDFT }&\textbf{DMC} &\textbf{LDA}&\textbf{PBE}&\textbf{PBEsol} & \textbf{SCAN}\\
                \textbf{}&\textbf{($erg/cm^{2}$)}&\textbf{} &\textbf{}&\textbf{}&\textbf{} & \textbf{}\\
                2 & 3446 & $3392\pm 50$ & -3.2 & -5.8 & -2.7 & -0.7\\ 
                3 & 797 & $768\pm10$ & -4.1 & -7.0 & -2.9 & -1.1\\
                4 & 278 & $261\pm8$ & -4.1 & -9.4 & -4.0 & -1.6 \\
                6 & 58 & 53       & -6.1  &  -10.3  & -2.9 & 1.6 \\
            \end{tabular}
            \hfill{}
            \caption{
                Exchange-correlation energies ($[{{\sigma}_{XC}}-{{\sigma}_{Exact-XC}}]/{{\sigma}_{Exact-XC}}$) 
                of the jellium surface for different values of $r_{s}$.
            }
            \label{tab:exc_jellium} 
            \end{ruledtabular}
        \end{table}

   \begin{table}[H]
       \begin{ruledtabular}
            \begin{tabular}{ccccc}
                \textbf{Error}&\textbf{LDA}&\textbf{PBE}&\textbf{PBEsol} & \textbf{SCAN}\\
                \textbf{}&\textbf{}&\textbf{}&\textbf{} & \textbf{}\\
                ME (Ha)      &   -41.38 & -72.23 & -32.37 & -9.14\\
                MARE (Ha)    &   41.38 & 72.23 & 32.37 & 9.60\\
                MRE (\%) &   -5.5 & -20.93 & -3.13 & -0.45\\
                MARE (\%) &   5.5 & -20.93 & 3.13 &  1.25 \\
                RMSD (\%) &   2.4 & 17.61 & 0.59 & 1.42\\
            \end{tabular}
           \hfill{}
            \caption{
                Error in Exchange-correlation energies of jellium surface for different values 
                of $r_{s}$ calculated for the values in Table~\ref{tab:exc_jellium}.
            }
            \label{tab:error_exc_jellium} 
            \end{ruledtabular}
        \end{table}
    
    \subsection{Computational details:}
    \label{appendix:computational-details}
    We performed first-principles density functional theory (DFT) calculations
    using the VASP package \cite{hafner2008ab} in combination with projector augmented wave (PAW)
    method \cite{blochl1994projector,KJ99}. For both bulk and surface computations, a maximum kinetic energy
    cutoff of 700 eV was used for the plane-wave expansion. The Brillouin zone was sampled
    using Gamma centered k-mesh grids of size $16 \times 16 \times 16$ for the bulk and $16 \times
    16 \times 1$ for the surfaces. The top few layers in the slab were fully relaxed until
    the energy and forces were converged to 0.001 eV and 0.02 eV/{\AA}, respectively.
    Dipole corrections were employed to cancel the errors of the electrostatic potential,
    atomic forces and total energy, caused by periodic boundary condition.
 
    For the slab geometry, 20{\AA} of vacuum was used to reduce the Coulombic interaction
    between the actual surface and its periodic image. For (111) surfaces, a hexagonal
    cell was used with each layer containing one atom per layer. The same procedure
    was employed for (100) tetragonal and (100) orthorhombic cell. The cells are built
    using the theoretical lattice constants obtained from fitting the Birch-Murnaghan(BM)
    equation of state for the bulk with each functional, see Tab.~\ref{tab:lattice-constant}. 
    We do not consider exchange-correlation contributions to the planar averaged local electrostatic potential ($V_{Vacuum}$). We used Pt (111) to test the convergence of the surface properties with respect to different computational variables such as k-mesh, cut-off energy, layer and vacuum thickness of the slab geometry. All the computed surface properties presented in this work, are well converged with respect to these computational variables. 
         \begin{table*}[ht]
        \begin{ruledtabular}
            \begin{tabular}{ccccccc}
                \toprule
                \textbf{Metals}&\textbf{LDA}&\textbf{PBE}&\textbf{PBEsol}&\textbf{SCAN} &
                \textbf{SCAN+rVV10}  & \textbf{Experimental}  \\   
                \midrule
                Al &  3.981 & 4.034 & 4.008 &    4.004 & 3.996    & 4.018   \\ [0.2cm]       
                Cu &  3.524 & 3.631 & 3.561 & 3.558 & 3.545    & 3.595  \\ [0.2cm]         
                Ru & c=4.265 & c=4.269 & c=4.267 & c=4.265 & c=4.266  & c=4.281  \\
                [0.2cm]   
                   & c/a =1.571 & c/a=1.574 & c/a=1.573 & c/a=1.572 & c/a=1.575 & c/a= 1.582 \\
                [0.2cm]
                Rh &  3.751 & 3.825 & 3.8 & 3.784 & 3.773     &  3.794  \\ [0.2cm]          
                Pd &  3.834 & 3.935 & 3.866 & 3.896 &  3.877    & 3.876   \\ [0.2cm]       
                Ag &  4.001 & 4.145  & 4.079  & 4.087 & 4.060     &  4.062  \\ [0.2cm]        
                Pt &  3.897  & 3.967 &  & 3.919 & 3.896    & 3.888     \\ [0.2cm]     
                Au &  4.052  & 4.156  & 4.079  & 4.087 & 4.073    & 4.063    \\ [0.2cm]      
            \end{tabular}
        \end{ruledtabular}
    
        \caption{
            Calculated values of the lattice constants of the metals. Experimental data are taken form ref\onlinecite{KressePRB2010,KressePRB2013}.
        }
        \label{tab:lattice-constant}
    \end{table*}

    \bibliography{prxref}{}

\end{document}